\documentstyle[aps,pre,epsf,multicol]{revtex}

\begin{document}
\draft
\title{Stationary State Skewness in Two Dimensional KPZ Type Growth.}
\author{ Chen-Shan Chin and Marcel den Nijs}
\address{ Department of Physics, University of Washington, P.O. Box 351560, \\ 
Seattle, Washington 98195-1560}
\maketitle

\begin{abstract}
We present numerical Monte Carlo results for the stationary state properties 
of KPZ type growth in two  dimensional surfaces, 
by evaluating the finite size scaling (FSS) behaviour of 
the 2nd and 4th moments, $W_2$ and $W_4$, 
and  the skewness, $W_3$, in the Kim-Kosterlitz (KK)  and BCSOS model.
Our results agree with the stationary state proposed by L\"assig.
The roughness exponents $W_n\sim L^{\alpha_n}$ obey power counting,
$\alpha_n= n \alpha$, and the amplitude ratio's of the moments are universal.
They have the same values in both models:
$W_3/W_2^{1.5}= -0.27~(1)$ and $W_4/W_2^{2}= +3.15~(2)$.
Unlike in one dimension, the stationary state skewness is not tunable,
but a universal property of the stationary state distribution.
The FSS corrections to scaling in the KK model are weak and 
$\alpha$ converges well to the Kim-Kosterlitz-L\"assig value $\alpha=\frac{2}{5}$.
The FSS corrections to scaling in the BCSOS model are strong.
Naive extrapolations yield an smaller value, $\alpha\simeq 0.38 (1)$,
but are still consistent with $\alpha=\frac{2}{5}$ if the leading irrelevant
corrections to FSS scaling exponent is of order $y_{ir}\simeq -0.6~(2)$.
\end{abstract}

\begin{multicols}{2}
\section{Introduction}\label{sec:1}

KPZ type growth is one of the generic dynamic processes
describing growth of crystal surfaces.
It is named after the Langevin equation introduced by 
Karder, Parisi, and Zhang about a decade ago ~\cite{KPZ,rev1,rev2,rev3,rev4}.
\begin{equation}
\label{KPZ-eq}
\frac{\partial h}{\partial t} = 
\nu \frac{\partial h^2}{\partial^2 x} +
\lambda  \left(\frac{\partial h}{\partial x}\right)^2 + 
\eta  
\end{equation}
with $\eta$ uncorrelated noise 
\begin{equation}
\label{noise-eq}
\left<\eta(x_1,t_1) \eta(x_2,t_2) \right>=  \rm{D} \delta(t_1-t_2) \delta(x_1-x_2)  
\end{equation}
Numerous microscopic models on the master equation level
have been studied numerically as well and are confirmed to be in the KPZ 
universality class~\cite{rev1,rev2,rev3,rev4}.
However, many properties of this process are still in question,
including basic aspects, like the precise values of the 
scaling critical exponents,
and the detailed structure of the stationary growing state.
Part of the problem is the absence of an obvious  mean field theory.
The linear, integrable diffusion, part of the KPZ equation ($\lambda=0$)
does  not play the role of mean field theory fixed point 
in high enough dimensions ($D$). The KPZ behaviour is governed by a 
strong coupling fixed point for all $D$, 
and thus evades perturbative renormalization treatments~\cite{KPZ-mf}.

It is widely accepted that the dynamic exponent $z$ and the stationary state 
roughness exponent $\alpha$ obey the equality $\alpha+z =2$ 
in all $D$~\cite{rev1,rev2,rev3,rev4}.
These critical exponents specify how time and height rescale under
a renormalization transformation: 
$x \to b x$,  $t \to b^z t$, and  $h \to b^\alpha h$.
The exponent identity states that under renormalization the 
amplitude of the non-linear term in the KPZ equation does not change.
$\lambda$ is a so-called redundant scaling field.
It plays a  role similar to lattice anisotropy in equilibrium phase transitions. 
Increasing $\lambda$ simply speeds-up the process, 
and scaling amplitudes are proportional to $\lambda$. 
This exponent equality links the dynamic scaling to the stationary state scaling.
Therefore the focus has shifted recently 
to the structure of the stationary state.

The stationary growing  state is trivial in one dimension ($1D$). 
It is the Gaussian distribution. 
The up and down steps along the surface are uncorrelated beyond
a definite correlation length. 
This implies $\alpha=\frac{1}{2}$ (the random walk value), 
and from the above exponent identity it follows that  $z=\frac{3}{2}$. 
This behaviour is well established, 
not only by numerical studies~\cite{rev1,rev2,rev3,rev4}, 
but also analytically. 
The $1D$ body centered solid on solid (BCSOS) growth model is  
exactly soluble~\cite{Dhar,Spohn}.
It's master equation is a special case of the 2D equilibrium 6-vertex model. 
In the latter representation, KPZ scaling describes facet ridge end-points 
of equilibrium crystal shapes~\cite{JN-MdN}.
$1D$ KPZ growth is equivalent also to  
asymmetric exclusion hopping processes~\cite{Derrida}. 
Moreover, the exact stationary state of the 
Langevin equation itself is known in $1D$
and is indeed the Gaussian distribution~\cite{rev4}).

The stationary state is not simple in $D>1$.
The stationary state roughness exponent $\alpha$ takes a non-trivial value
and is actually not very well known numerically.
For example, in $2D$ the reported values vary between $\alpha=0.37-0.4$
~\cite{rev2,rev3,KimKos,KK-temp,BCSOS-num}.
L\"assig made an important analytical break through last year~\cite{Lassig}. 
He proposed the likely structure of the stationary state by studying the operator 
product expansion of the (height variable) correlation functions. 
Under the assumption that the algebra closes and contains only one
scaling field operator, L\"assig obtained a quantization condition for 
the exponents. One of these solutions, $\alpha=\frac{2}{5}$, 
is close to the above 2D numerical values. 
The moments 
\begin{equation}
\label{moments}
W_n = \left< \left( h_i -\bar h \right) ^n \right>
\end{equation}
of  L\"assig's stationary state distribution obey power counting, 
i.e., the exponents in the scaling relations
\begin{equation}\label{moments-scaling}
W_n( N^{-1},t^{-1}) = b^{\alpha_n}  W_n( b N^{-1},b^z t^{-1}).
\end{equation}
are related as $\alpha_n=n \alpha$.
The distribution lacks multi scaling. 
Moreover, for the closure of the algebra it is important that
the stationary state is skewed.
The odd moments, in particular the third one must be non-zero.

In this paper we report a detailed numerical study of the stationary 
state in 2D for the Kim-Kosterlitz (KK)~\cite{KimKos} and the BCSOS model. 
We determine the finite size scaling behaviour of the second, third, 
and fourth moments. 

In section 2 we review the properties of   
stationary state skewness in $1D$ KPZ growth, 
and list the possibilities in higher dimensions.
Section 3 contains our numerical results for the KK and BCSOS model. 
The third moment is indeed non-zero and  
the second, third, and fourth moments obey indeed power counting. 
We find strong evidence that the amplitude ratio's 
of the moments are universal, 
$R_3= W_3/W_2^{1.5}= -0.27 (1)$ and 
$R_4= W_4/W_2^{2}  = +3.15 (2)$.
The amount of stationary state skewness is the same in these two models.
In $1D$ skewness is tunable, but appears not to be so in $2D$.
The values of these amplitude ratio's are the major new results of this paper. 

In section 3 we address also whether the difference in the previously reported values 
of $\alpha$ for the KK and BCSOS models can be attributed to 
finite size scaling  (FSS) type corrections.
The FSS corrections to scaling exponent is large, 
but in the same range of values suggested by naive power counting.
  
To check more directly whether skewness is tunable or not,
we introduce a temperature type parameter $K$ in the BCSOS model.
In the KK model it is known that $\lambda$ changes sign with $K$~\cite{KK-temp}.
In section  4 we give an intuitive explanation for why this happens.
(It is related to preroughening phenomena in equilibrium surfaces.)
In the BCSOS model $\lambda$ does not change sign. 
In section 5 we present MC data for the $K$ dependence of the
roughness exponents and the amplitude ratio's.
Both show some systematic drift, but much  
smaller than expected  if they would vary with $K$.   

\section{Stationary State Skewness}\label{sec:2}

This study was actually not motivated by  L\"assig's recent results.
It was conceived as a generalization of an earlier study
of stationary state skewness in $1D$~\cite{JN-MdN-sk}.
The KK model is a special point in the restricted solid-on-solid (RSOS) model. 
We varied its adsorption and evaporation probabilities in the 1D model
by making them dependent on the local nearest neighbour heights.
That lead to 5 independent parameters. 
Surprisingly we were able to construct the exact stationary state in a 
4 dimensional subspace. Its structure is simple.
The steps in the interface are completely uncorrelated.
Only the step density varies. This state has zero skewness. 
It is a Gaussian and has particle-hole symmetry.
Outside this exact soluble subspace the stationary state is skewed. 
This means that in general KPZ type growth in $1D$ has a non-zero third moment 
$W_3$ in its stationary state.
For example, the KK point lies outside the non-skewed subspace.
On the other hand, the stationary states in the 
exactly soluble BCSOS model and also the 
Langevin equation itself are non-skewed.
Stationary state skewness is distinct from temporal skewness. 
Most initial states develop into skewed  structures at intermediate time scales 
(temporal skewness) even if the stationary state is not skewed~\cite{temp-sk}.

So in $1D$, KPZ type stationary states are typically skewed, and its amplitude
is tunable. This raises the immediate question whether skewness 
affects the scaling exponents.
Let's refer to the operators leading out of the non-skewed subspace as 
${\cal O}_{sk}$ and to their conjugate coupling constants as $u_{sk}$.
In $1D$, a KPZ type fixed point with  zero-skewness exists.
The question is whether this fixed point is stable; whether 
${\cal O}_{sk}$ is a relevant, marginal, redundant, or an irrelevant operator.
Suppose the non-skewed KPZ fixed point is stable.
The moments should scale then with system size as
\begin{equation}
\label{moments-sk}
W_n( N^{-1}, u_{sk}) = b^{\alpha_n} W_n( bN^{-1}, b^{y_{sk}} u_{sk}) 
\end{equation}
with $\alpha_n= n \alpha$ and $y_{sk}<0$.
All even moments scale as 
\begin{equation}
W_n\simeq A N^{\alpha_n}+.... 
\end{equation}
with universal amplitude ratio's, $R_n=W_n/W_{2}^{n/2}$.
All odd moments scale as
\begin{eqnarray}
\label{W-fss}
W_n( N^{-1}, u_{sk}) & = & N^{\alpha_n}  {\cal F}_n( N^{y_{sk}} u_{sk}) \\ \nonumber
                     & = & N^{\alpha_n} \left[{\cal F}_n( 0) +{\cal F}_n^\prime(0) N^{y_{sk}} u_{sk}+...\right] \\ \nonumber
                     &\sim& u_{sk} N^{\alpha_n+y_{sk}}
\end{eqnarray}
with ${\cal F}_n( 0)=0$ because the fixed point has no skewness.
The odd amplitude ratio's are proportional to $u_{sk}$,
i.e., the skewness varies continuously.

Numerical (transfer matrix finite size scaling) results confirmed 
that the non-skewed KPZ fixed point is stable in $1D$. 
We found $y_{sk}\simeq -1$.
Moreover, the amplitude of $W_3$  (at e.g., the KK point) is indeed 
roughly proportional to the skewness coupling constant $u_{sk}$,
in accordance with eq.\ref{W-fss}.

Skewness is negative at the $1D$  KK point. 
On average hill tops are wider (flatter, less sharp) than valley bottoms.
Such a statement is meaningless without a specification of a cut-off.
The definition of what constitutes a mountain and what represents a
local hump depends on the length scale at which 
the surface configuration is being viewed.
(Humans do not interpret every grain of sand as a hill.)
Skewness is a scale dependent property.
The asymptotic scaling of the moments tells us how asymmetric
the hills and valleys are in the large length scale limit.
In $1D$ this skewness persists  all the way 
to microscopic length scales.
We calculated the surplus of sharp valley bottoms over sharp hill tops
at the  microscopic (the grains of sand) level.
This expectation value has the same sign as the macroscopic skewness  
and is also  roughly proportional to $u_{sk}$.
This invariance of the surface structure over all length scales 
is related to the rather trivial nature of the 
fixed point stationary state (Gaussian distribution).

The tunability of the skewness at the microscopic length scale is
easy to understand. Consider the $1D$ RSOS model with deposition only,
with three parameters: $p_h$, $p_s$, and $p_v$~\cite{JN-MdN-sk}.
The density of local sharp hill tops is set by the deposition probability 
$p_h$ of particles on local flat surface segments.
The sharpness of these local hilltops is set by the rate at which they broaden,
i.e., the probability with which particles adhere to existing steps, $p_s$.
The density of local sharp valleys is set by 
the rate at which single particle puddles fill-up, $p_v$  compared to 
the rate at which they  are created, $2p_s$.
These processes balance exactly inside the subspace where 
the stationary state is trivial and has zero skewness.
At the KK point, $p_h=p_s=p_v=1$, the balance is imperfect and 
the dynamics creates a backlog of ``to-be-filled-up" local valleys. 
Newly created local hill tops broaden readily, 
and therefore are flatter~\cite{JN-MdN-sk}.

Is the skewness tunable in $D>1$ as well?
Is there maybe a line of KPZ type fixed points with continuously varying skewness?
A varying exponent $\alpha$  would explain the current numerical spread in its value.
Is it possible to change the sign of skewness without
changing the sign of $\lambda$ in eq.(\ref{KPZ-eq})?
Or. is there one single KPZ stationary state fixed point distribution, 
with a universal amount (and sign) of skewness?
In that case we need to explain the numerical spread in $\alpha's$
in terms of strong FSS corrections.
These are the issues we address in this paper.

\section{Scaling of the Stationary State Moments}\label{sec:3}

We perform a systematic numerical study of the stationary state properties 
in the $2D$ KK-model and the BCSOS model.
In both cases we  allow only particle deposition (no desorption).
Consider a $2D$ square lattice with an height variable 
$h(r)= 0, \pm1,\pm2,...$ at each lattice site.
We apply periodic boundary conditions.
In the KK-model, nearest neighbour columns are allowed to differ by at most 
one unit, $\delta h=0,\pm1$.
Choose a site at random, and deposit a particle,
$h \to h+1$, with probability $p=1$  unless such a move would violate the 
above restriction.

In the BCSOS model the square lattice is divided into two 
sublattices. The height variables are restricted 
to be even on one of them, $h(r)= 0, \pm2,...$, 
and to be odd on the other $h(r)= \pm1,\pm3, ...$, 
such that nearest neighbour columns always
differ in height by one $\delta h=\pm1$.
Choose at random a site, and deposit a particle with probability $p=1$
if the move does not violate the $\delta h=\pm1$ constraint.
In section \ref{sec:5} we consider also the generalization 
where the BCSOS deposition probabilities
vary with the local configuration by means of 
a temperature type parameter, $p= {\rm min}(1, \exp(-\Delta E))$.
$\Delta E$ is the energy change if the adsorption would take place.
The energy 
\begin{equation}
\label{BCSOS}
E(\{h_i\})= \sum_{\left< i,j\right>}\frac{1}{4} K (h_i-h_j)^2 
\end{equation}
has a tunable parameter $K$ and  
the summation runs over all next nearest neighbours.
This rule is identical to that in standard Metropolis MC simulations,
except that desorption is forbidden.
The latter breaks detailed balance and leads to the 
non-equilibrium growing  stationary state.

We determine the second, third, and fourth moments of the stationary states.
The MC averages in this section involve  $\simeq 2~10^6$ MC steps,
after  $\simeq 4~10^3$ initial MC configurations, 
to allow the surface to reach its stationary state.
The square lattice  size $L^2$ varies between  $12\leq L\leq128$. 

\begin{figure}
\centerline{\epsfxsize=8cm \epsfbox{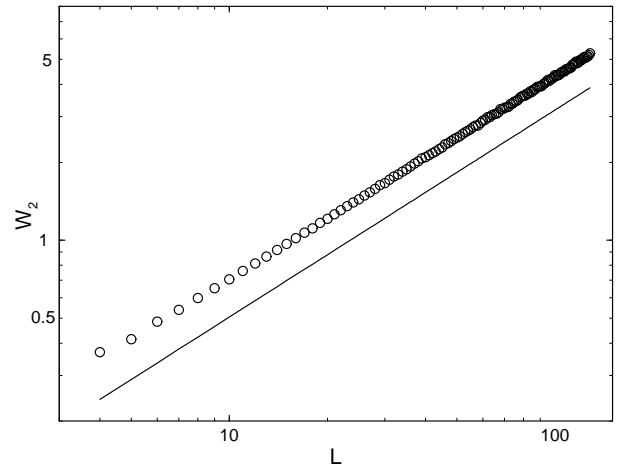}}
\caption{ Stationary state second moment $W_2$ 
of the 2D Kim-Kosterlitz (KK) model as function of lattice size $L^2$.
The solid line with slope 0.8 is shown as reference.}
\label{RSOS-W2}
\end{figure}

\begin{figure}
\centerline{\epsfxsize=8cm \epsfbox{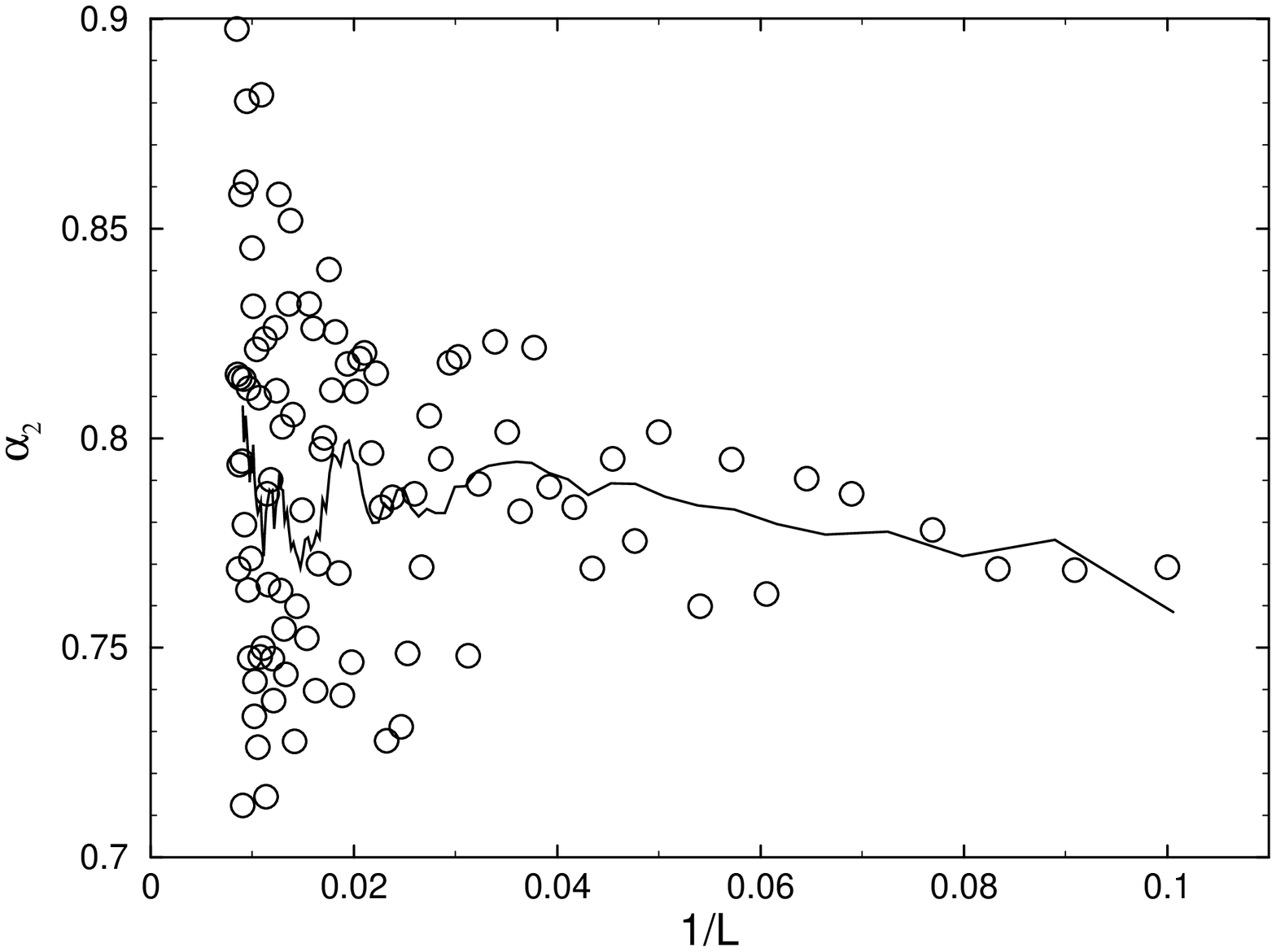}}
\caption{ 
Finite size scaling approximants for the surface roughness exponent 
$\alpha_2$ of the second moment, $W_2\simeq A L^{\alpha_2}$, 
in the stationary state of the 2D KK model.}
\label{RSOS-a2}
\end{figure}

First consider the KK model.
Fig.~\ref{RSOS-W2} shows the second moment $W_2$.
It scales as $W_2\simeq A L^{\alpha_2}$. 
The slope of the log-log plot gives the exponent $\alpha_2$.
It is a mistake to apply a least-square type fit to the slope at large $L$.
One should determine the slope at various system size intervals and
perform a finite size scaling (FSS) analysis.
Fig.~\ref{RSOS-a2} represents such an analysis.
It  shows FSS estimates for $\alpha_2$
from the same data, defined as
$\alpha_2(L) = \log[W_2(L_2)/W_2(L_1)] / \log[L_2/L_1]$
with $L_2\simeq1.2 L_1$ and $L=\frac{1}{2}(L_1+L_2)$.
Such a FSS scaling analysis is only barely feasible 
at large system sizes due to the intrinsic MC noise.
Fig.~\ref{RSOS-a2} is indeed rather noisy at large $L$. 
The MC scatter increases with system size, since 
the stationary state is intrinsically critical and therefore 
subject to critical slowing down.
We opted for running many lattice sizes instead of fewer but longer MC runs. 
The solid line is obtained by averaging the $\alpha_2(L)$ locally, 
over $L-7\leq L\leq L+7$.
The data in Fig.~\ref{RSOS-a2} converge by eye to $\alpha_2=0.80(2)$, 
consistent with the value $\alpha=\frac{2}{5}$ proposed by Kim and Kosterlitz
from their earlier numerical results~\cite{KimKos,KK-temp},
and with  L\"assig's~\cite{Lassig} stationary state.
The FSS corrections to scaling in the KK model are small compared to the
MC noise.

\begin{figure}
\centerline{\epsfxsize=8cm \epsfbox{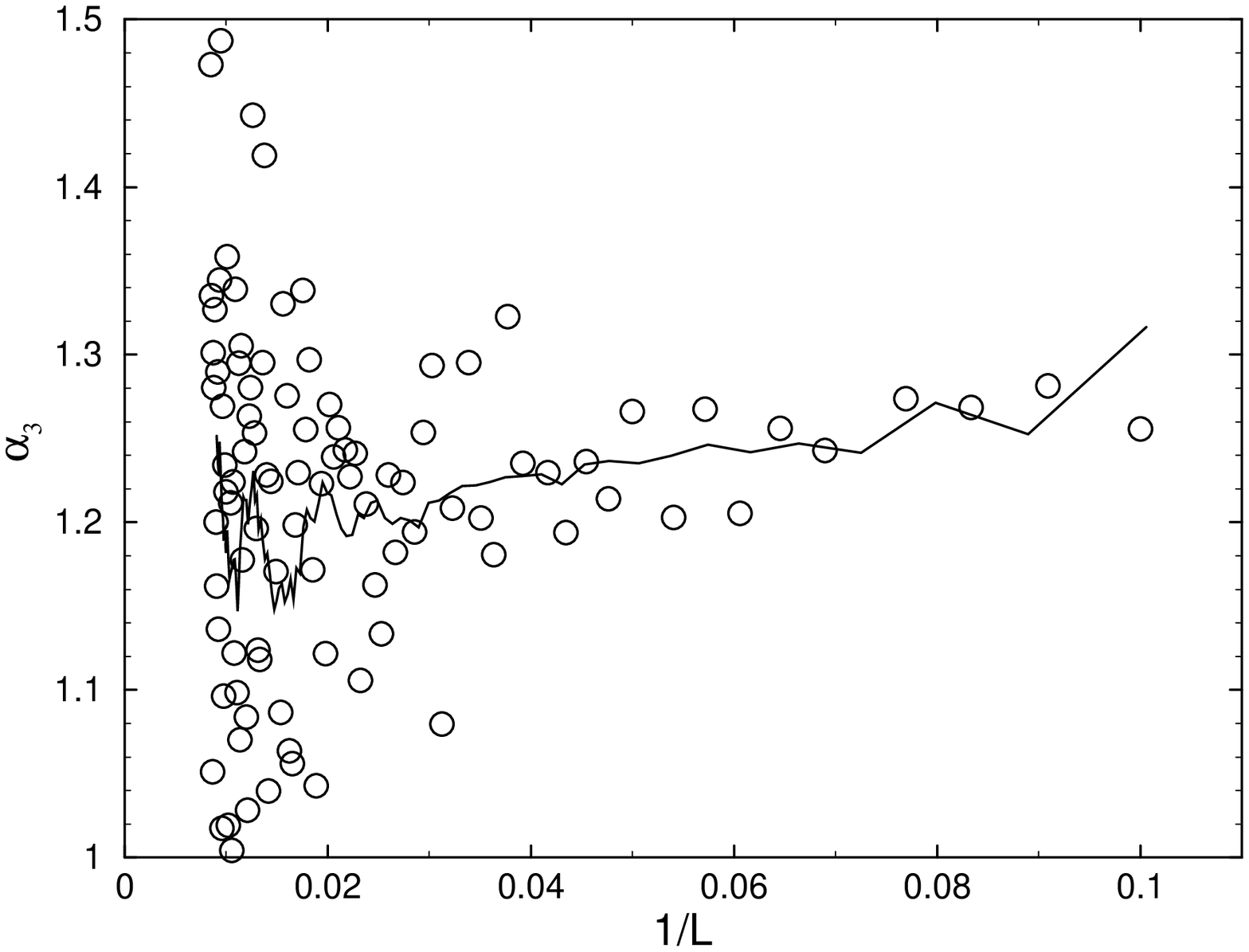}}
\caption{ 
Finite size scaling approximants $\alpha_3(L)$ for the surface roughness exponent  
$\alpha_3$ of the third moment, $W_3\simeq A L^{\alpha_3}$, 
in the stationary state of the 2D KK model.}
\label{RSOS-a3}
\end{figure}

\begin{figure}
\centerline{\epsfxsize=8cm \epsfbox{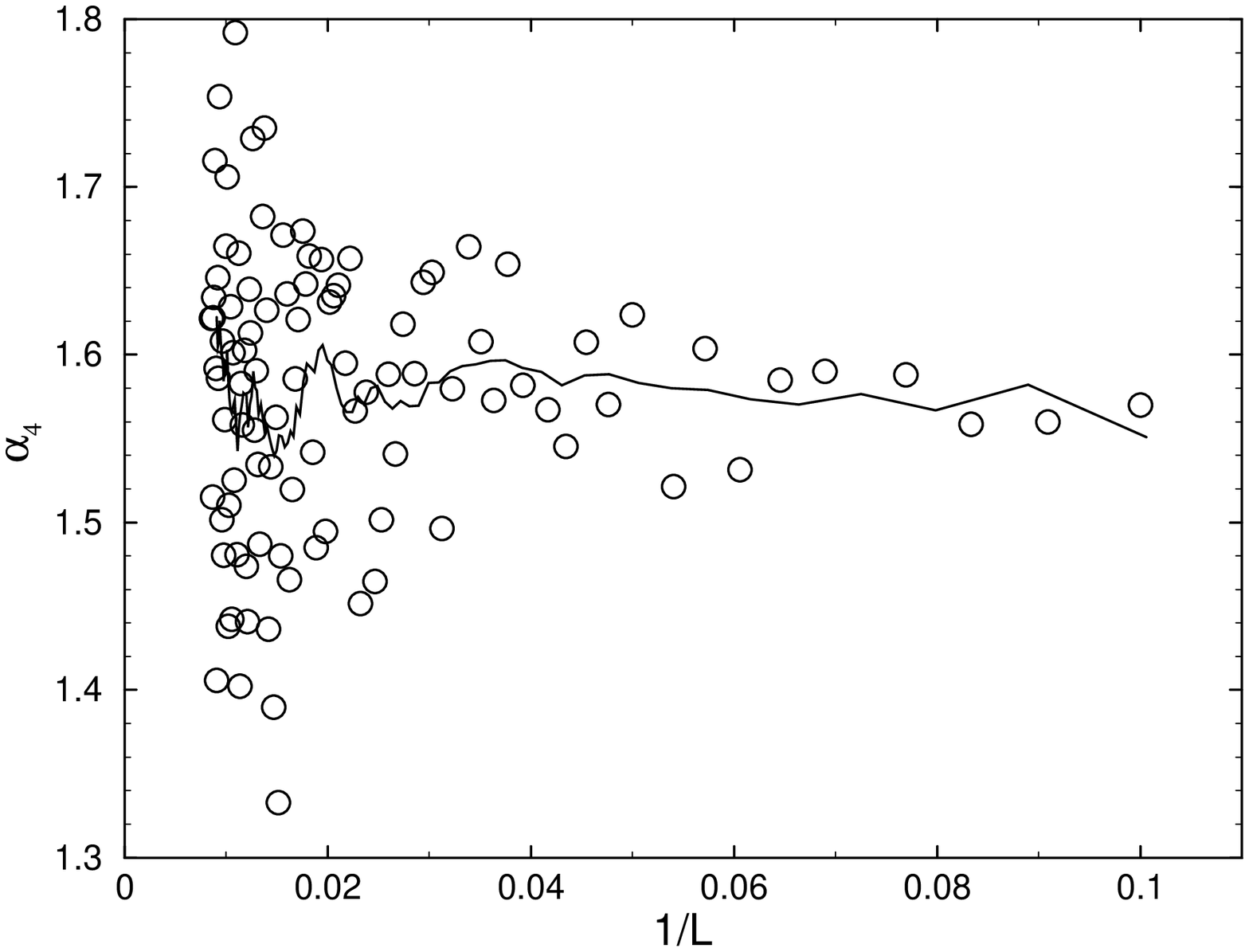}}
\caption{ 
Finite size scaling approximants $\alpha_4(L)$ for the surface roughness exponent  
$\alpha_4$ of the fourth moment, $W_4\simeq A L^{\alpha_4}$, 
in the stationary state of the 2D KK model.}
\label{RSOS-a4}
\end{figure}

Figs.~\ref{RSOS-a3} and \ref{RSOS-a4} show the same FSS analysis 
for the  third and fourth moment exponents $\alpha_3$ and $\alpha_4$.
They converge by eye to $\alpha_3= 1.20 (4)$  and $\alpha_4= 1.60 (5)$.
This agrees with L\"assig's stationary state.
In particular, the surface is skewed, and power counting, 
$\alpha_n=n\alpha$, is satisfied within the numerical accuracy.
The corrections to FSS are again small compared to the MC noise.

\begin{figure}
\centerline{\epsfxsize=8cm \epsfbox{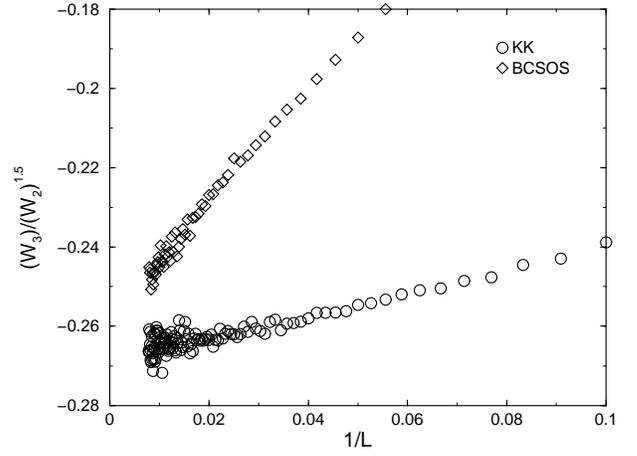}}
\caption{ 
Finite size scaling behaviour of the skewness amplitude ratio $R_3=W_3/W_2^{1.5}$ 
in the 2D KK model (the circles) and the $K=0$ BCSOS model (diamonds).}
\label{R3}
\end{figure}

\begin{figure}
\centerline{\epsfxsize=8cm \epsfbox{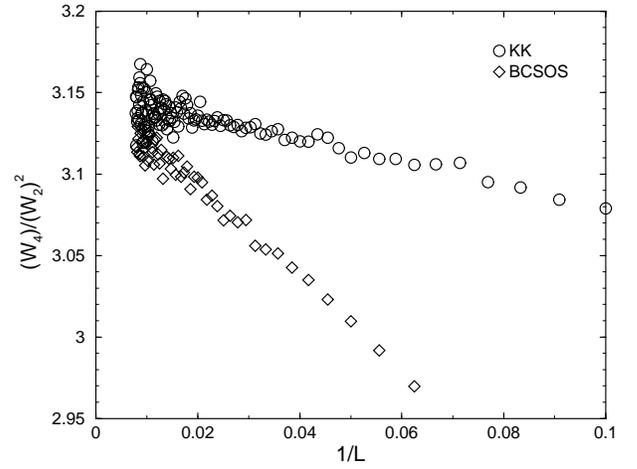}}
\caption{ 
Finite size scaling behaviour of the fourth moment amplitude ratio $R_4=W_4/W_2^2$ 
in the 2D KK model (the circles) and the $K=0$ BCSOS model (diamonds).}
\label{R4}
\end{figure}

Figs.~\ref{R3} and \ref{R4} show the amplitude
ratio's (the circles) of the third and fourth moment compared to the 
second one, $R_n= W_n/(W_2)^{n/2}$. 
The fact that these ratio's  convergence gives additional evidence of the 
validity of power counting.
Actually, they do so much smoother than the exponents $\alpha_n$, 
suggesting that the MC fluctuations tend to 
preserve the power counting property better than the precise value of $\alpha$.     
The amplitude ratio's converge smoothly to 
$R_3= -0.27 (1)$ and $R_4= +3.15 (2)$.
Skewness is negative, $R_3<0$.
It is probably wishful thinking to guess that $R_4=\pi$.
($R_4=3$ in Gaussian distributions.) 

\begin{figure}
\centerline{\epsfxsize=8cm \epsfbox{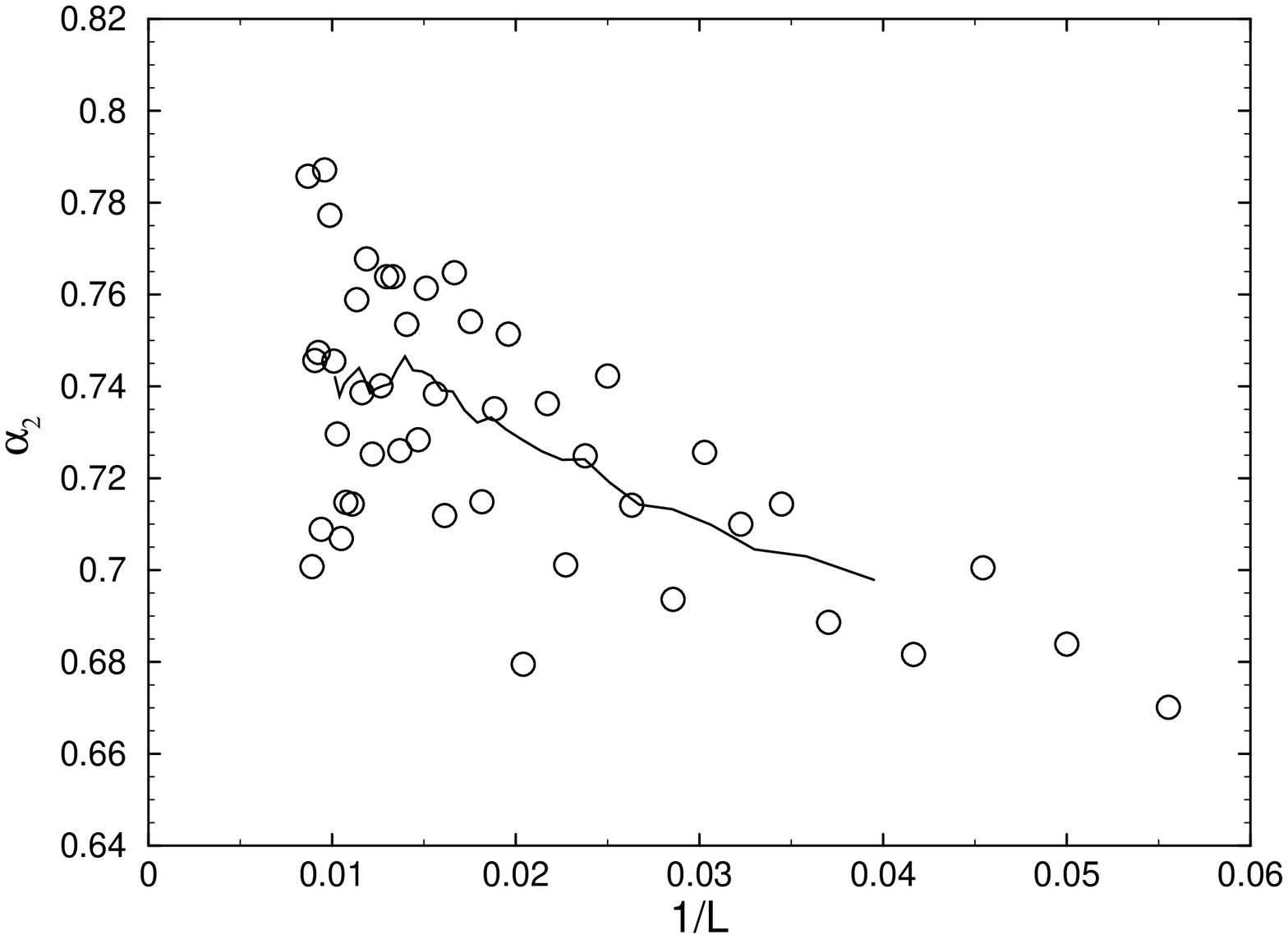}}
\caption{ 
Finite size scaling approximants for the surface roughness exponent  
$\alpha_2$ of the second moment, $W_2\simeq A L^{\alpha_2}$, 
in the stationary state of the 2D BCSOS model.}
\label{BCSOS-a2}
\end{figure}

\begin{figure}
\centerline{\epsfxsize=8cm \epsfbox{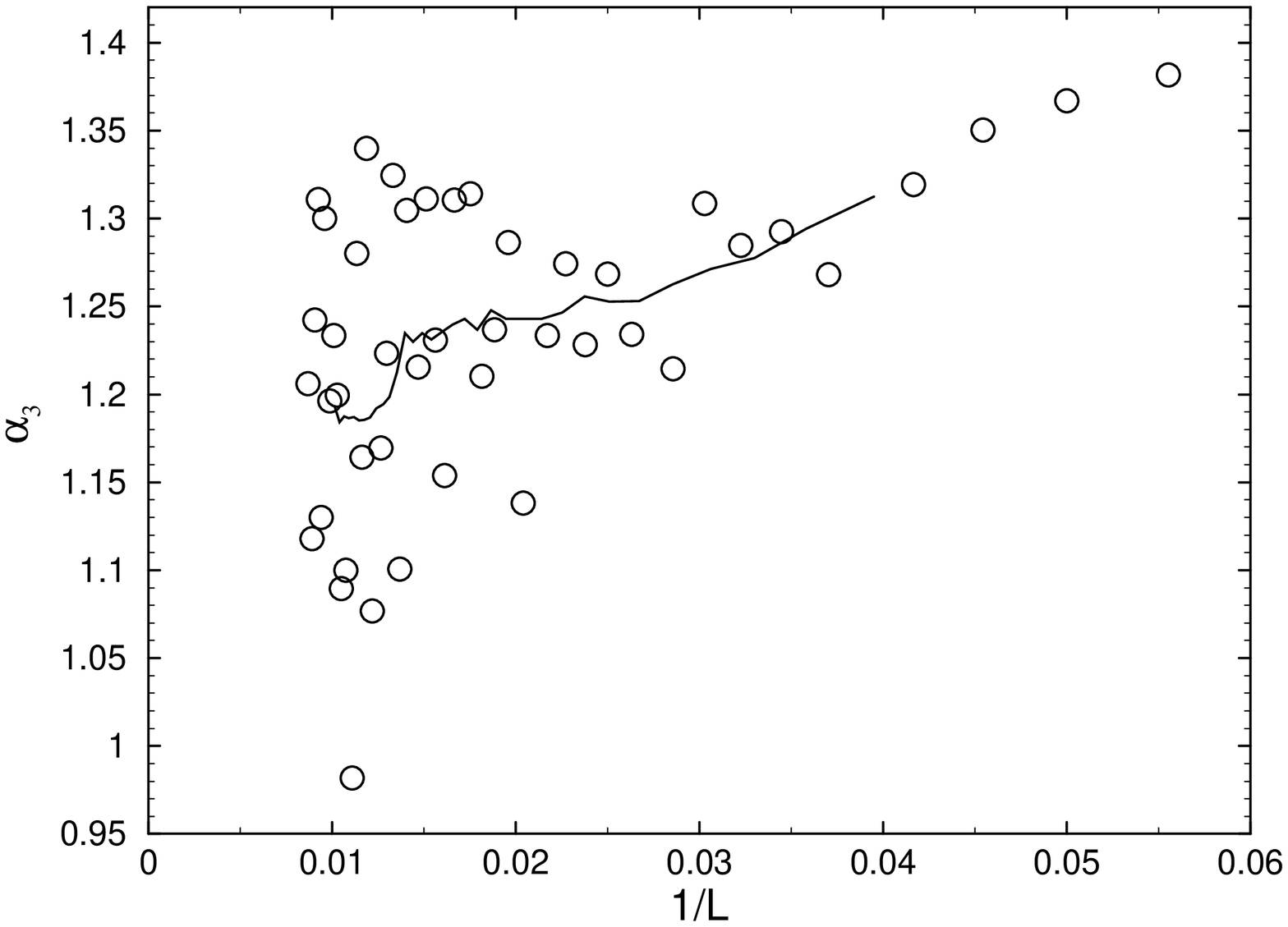}}
\caption{ 
Finite size scaling approximants for the surface roughness exponent 
$\alpha_3$ of the third moment, $W_3\simeq A L^{\alpha_3}$, 
in the stationary state of the 2D BCSOS model.}
\label{BCSOS-a3}
\end{figure}

\begin{figure}
\centerline{\epsfxsize=8cm \epsfbox{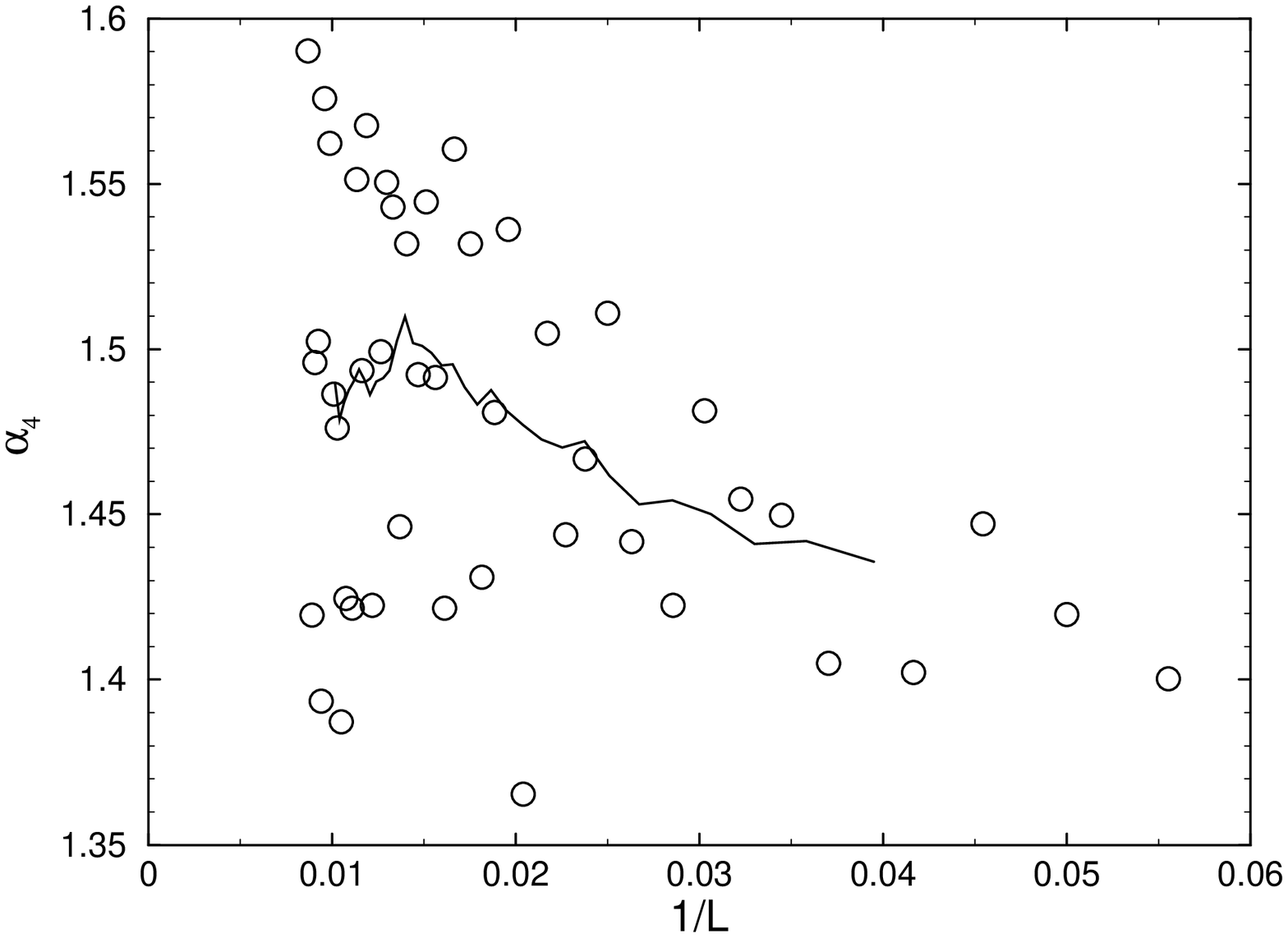}}
\caption{ 
Finite size scaling approximants for the surface roughness exponent 
$\alpha_4$ of the fourth moment, $W_4\simeq A L^{\alpha_4}$, 
in the stationary state of the 2D BCSOS model.}
\label{BCSOS-a4}
\end{figure}

Consider the BCSOS model at $K=0$. 
Figs.~\ref{BCSOS-a2}, \ref{BCSOS-a3}, and \ref{BCSOS-a4} 
show the same type of FSS estimates for the exponents $\alpha_n$. 
The finite size corrections to scaling are much larger than in the KK model. 
The data converge by eye systematically to smaller values
than in the KK model: 
$\alpha_2= 0.77 (2)$, $\alpha_3= 1.16 (3)$, and $\alpha_4 =1.54 (4)$.
This is consistent with the value for $\alpha_2$ reported in the 
literature~\cite{rev3,BCSOS-num}. 
These $\alpha_n$'s are mutually consistent with power counting.
Power counting is again more stable than the values of the exponents $\alpha_n$.
Most importantly, the amplitude ratio's in Figs.~\ref{R3} and \ref{R4} 
(the diamonds) converge to the same values as in the KK model (the circles).
 
Is $\alpha$ really smaller than in the KK model and
different from the Kim-Kosterlitz-L\"assig (KKL) value?
The finite size scaling corrections in the BCSOS model 
are several orders of magnitude bigger than in the KK model.
Is it believable that these apparent differences are due to 
corrections to scaling only?
The above FSS extrapolation ``by eye" presumes implicitly that all 
$\alpha_n$ converge approximately linearly in  $L^{-1}$.
This looks reasonable from the data, but is too restrictive.
Corrections to scaling originate from so-called irrelevant scaling fields. 
The corrections to scaling exponents $y_{ir}<0$
in $W_n \simeq A_n N^{\alpha_n} \left[ 1+ B_n L^{y_{ir}}+..... \right]$
are universal properties of the stationary state fixed point. 
The amplitudes $B_n$ are not universal. 
They depend on the ``distance" of the model to the fixed point
and the quantity we are looking at.
Assume that one correction to scaling term dominates,
i.e., that all other operators scale with much more negative values of $y_{ir}$.
The same exponent $y_{ir}$ should then appear in all moments. 
The only exception is that in specific quantities the leading term  might 
have zero amplitude by symmetry.
For example, all even moments might show a different leading exponent 
$y_{ir}$ than all odd ones. This actually happens here.
  
Suppose we force our BCSOS data to converge to $\alpha= \frac{2}{5}$.
We made plots of $W_n/L^{n \alpha}$ with $\alpha=\frac{2}{5}$ versus
$L^{y_{ir}}$ for a range of values of $y_{ir}$.
This should be a straight line at the proper $y_{ir}$.
From this we estimate that
$y_{ir}\simeq -0.6~(2)$ for the second and fourth moments, and
$y_{ir}\simeq -1.7~(3)$ for the third moment.
These straight line fits are satisfactory stable. 
So the KKL value for $\alpha$ is within the realm of possibilities
for the BCSOS model.  
Still, it remains a leap of faith, because the curves in 
Figs.~\ref{BCSOS-a2}, \ref{BCSOS-a3}, and \ref{BCSOS-a4}, 
are bend to the limit. 
 
It would be much more convincing if the corrections to scaling exponents
where known analytically, and/or take simple values.
At the core of L\"assig's result is the  assumption that the operator
content of the system is simply. Therefore one would expect that the
irrelevant operators have rather trivial critical dimensions,
like integers, multiples of $\alpha$, and combinations of both.
The above numerical values of $y_{ir}$ do not look that simple,
but are of the same order of magnitude as we might expect.
Simple minded power counting in Eq.(\ref{KPZ-eq}) with $\alpha=\frac{2}{5}$
and $z+ \alpha =2$, suggests that the corrections to scaling are strong; 
that the  curvature operator $\partial^2 h/ \partial x^2$ is irrelevant, 
but not by much, $y_{\nu}=-\alpha$, 
and that  the leading skewness operator 
$\left(\partial^2 h/ \partial x^2\right)^2$ scales with $y_{sk} =-2$.

\section{Temperature Dependent Transition Probabilities}\label{sec:4}

In section \ref{sec:5} we vary the temperature type parameter $K$ in the BCSOS model, 
to study the  universality of $\alpha$ and skewness issues 
raised in section \ref{sec:3} in more detail.
We do this only for the BCSOS model, not the RSOS model.
(The KK model is a special point in the latter.)
From earlier studies it is known that in the RSOS model
the KPZ non-linear term $\lambda$ changes sign with $K$~\cite{KK-temp}.
This creates strong Edwards-Wilkenson (EW) type 
corrections to scaling and will obscure the skewness property, because
the EW stationary state at $\lambda=0$ is non-skewed 
(the Gaussian distribution).
Such a change in $\lambda$ does not take place in the BCSOS model.

The following connection with preroughening phenomena in equilibrium 
crystal surface explains why $\lambda$  changes sign in the RSOS model
and not in the BCSOS model. 
Imagine a two dimensional phase diagram, with $K\sim T^{-1}$ 
the temperature like parameter, 
and a parameter $s$ representing the asymmetry between particle deposition 
and evaporation. $s=0$  corresponds to MC equilibrium type dynamics 
and  $s=1$ to the above pure deposition model without evaporation.
The equilibrium surface undergoes a roughening transition.
The rough phase is the EW stationary state.
The scaling properties of the equilibrium (stationary) state 
are described by the Gibbs distribution of the sine-Gordon model
\begin{equation}
\label{sine-Gordon}
E = \int dx~dy~ \left[ \frac{K}{2}  
(\nabla h)^2 + u_2 \cos(2 \pi h)+ u_4 \cos(4 \pi h) \right]
\end{equation}
It is known that $u_2$ varies with temperature and  changes sign inside the equilibrium 
rough phase in the RSOS model just above the roughening transition.
This follows from an exact duality transformation~\cite{MdN-RSOS}.
The location of this $u_2=0$ point at $s=0$
agrees qualitatively  with the numerical value
of the  $\lambda=0$ point at $s=1$.
Those values do not have to be  identical.
They only need to be of the same order of magnitude.
The following arguments suggest that a line of
$\lambda=0$ points emerges from the $u_2=0$ point at $s=0$ 
into the $s$ direction. 

$u_2$ is negative at the high temperature side of the $u_2=0$ line.
There, the  rough stationary state surface takes locally the 
so-called disordered flat type structure with alternating
up and down steps and local half integer surface height~\cite{MdN-PR}.
This is the same surface structure as in the BCSOS model
(but not  close-packed with $dh=\pm1$ kinks).
The non-linear term in the KPZ equation controls 
the local growth velocity at sloped sections of the surface. 
Growth at slopes is suppressed in BCSOS type rough structures, 
and therefore $\lambda<0$.
At the opposite, low temperature  $u_2>0$  side of the $u_2=0$ point,
the local rough RSOS surface is smooth, 
and has on the local level integer average surface heights.
In such structures, growth at sloped parts of the surface is 
enhanced and therefore $\lambda>0$.

The location of $u_2=0$  can be controlled  by introducing 
further neighbour interactions. 
This point transforms into the preroughening transition point
when it moves below the roughening temperature ~\cite{MdN-PR}.
(But only at $s\neq 0$, because the driven non-equilibrium surface is always rough). 

Such a  change in $\lambda$ does not take place in the BCSOS model. 
Its equilibrium stationary state  (at $s=0$) is exactly known for all $K$ 
from the exact solution of the 6-vertex model~\cite{6vertex}.
$u_2$ is negative at all values of $K$ in the $s=0$ equilibrium surface. 
Therefore there is no reason to expect a change of $\lambda$  as function 
of $K$ in the pure deposition model at $s=1$.
This makes the BCSOS model a suitable testing ground for the tunability
of $2D$ skewness and the universality of $\alpha$.

\section{Temperature Dependent Deposition Rates in the BCSOS Model}\label{sec:5}

In this section we study the universality of the stationary state roughness 
exponent $\alpha$ and the skewness by varying the deposition probabilities 
in the BCSOS model.
This should clarify whether the skewness is truly universal or a tunable parameter.
We can control the microscopic particle-hole asymmetry explicitly.

\begin{figure}
\centerline{\epsfxsize=8cm \epsfbox{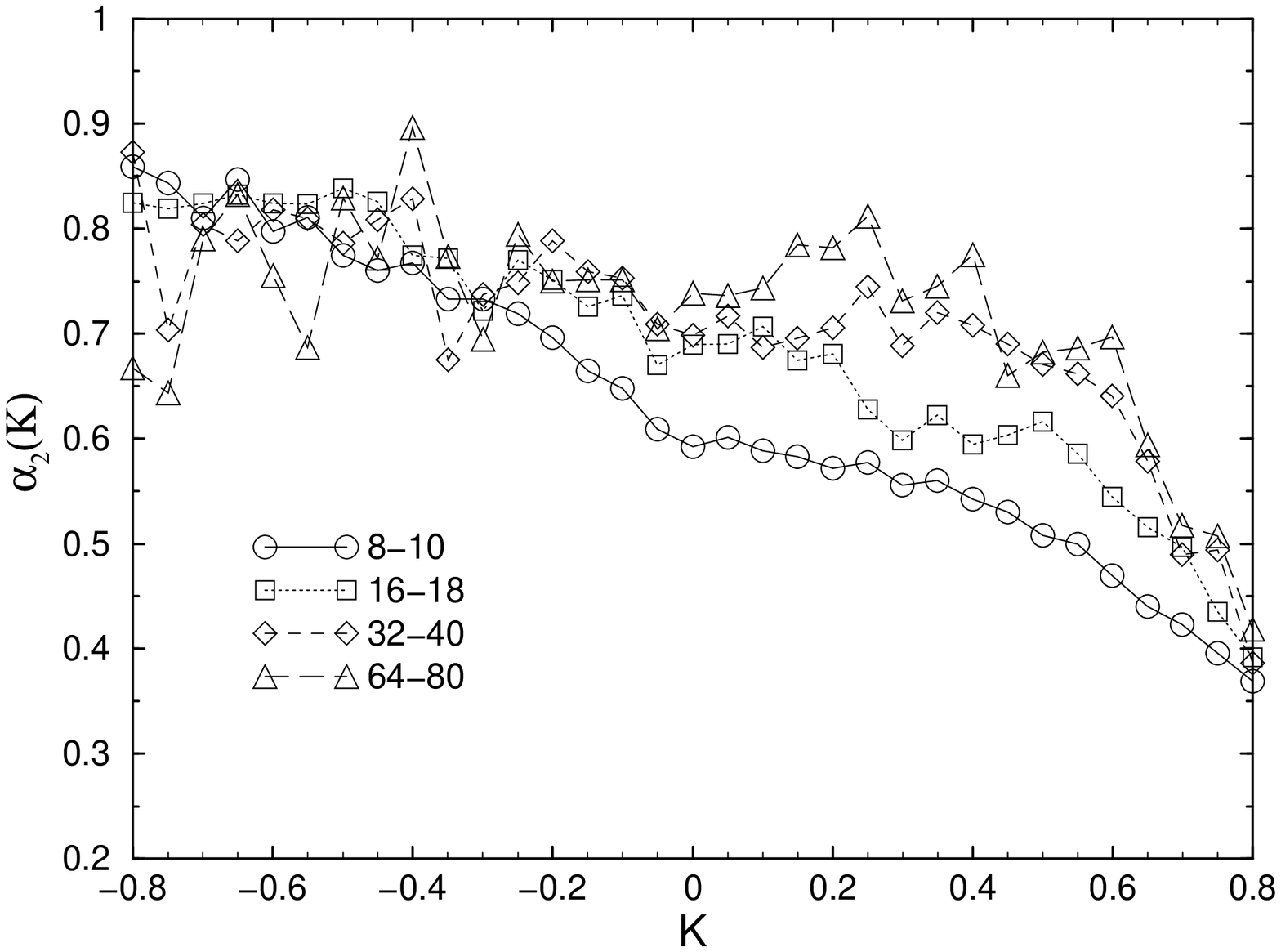}}
\caption{ 
Finite size scaling approximants for the surface roughness exponent 
$\alpha_2$ of the second moment, $W_2\simeq A L^{\alpha_2}$, 
in the stationary state of the 2D BCSOS model as function of $K$.}
\label{BCSOS-Ka2}
\end{figure}

 Fig.~\ref{BCSOS-Ka2} shows the variation of the second moment
exponent $\alpha_2(L)$ with the temperature parameter  $K$.
This plot is more qualitative than the ones in section \ref{sec:3}.
The system sizes are smaller (up to L=80) 
and the MC runs an order of magnitude shorter 
(up to $10^5$ MC steps, after 2500 initial MC steps). 
The curves have a definite slope at small system sizes, 
but these disappear with system size.
At temperatures beyond $K\simeq 1$ the surface becomes very flat and inactive. 
Compared to $K\simeq 0$ the system size is effectively much smaller and 
the MC runs effectively much shorter. 
This explains the decay in the approximants at large $K$.
At the opposite side, beyond $K\simeq -1$, the surface becomes
quite facetted at short distances and the dynamics slows down again.
Facetted structures have $\alpha =1$.
This explains why the $\alpha_2(L)$ curves drift upward 
on the left hand side in Fig.~\ref{BCSOS-Ka2}.
$\alpha_3$ and $\alpha_4$ vary also only weakly with $K$.
The amplitude ratios $R_3= W_3/W_2^{1.5}$ and $R_4=W_4/W_2^{2}$  
do not vary significantly with $K$ either. 

\begin{figure}
\centerline{\epsfxsize=8cm \epsfbox{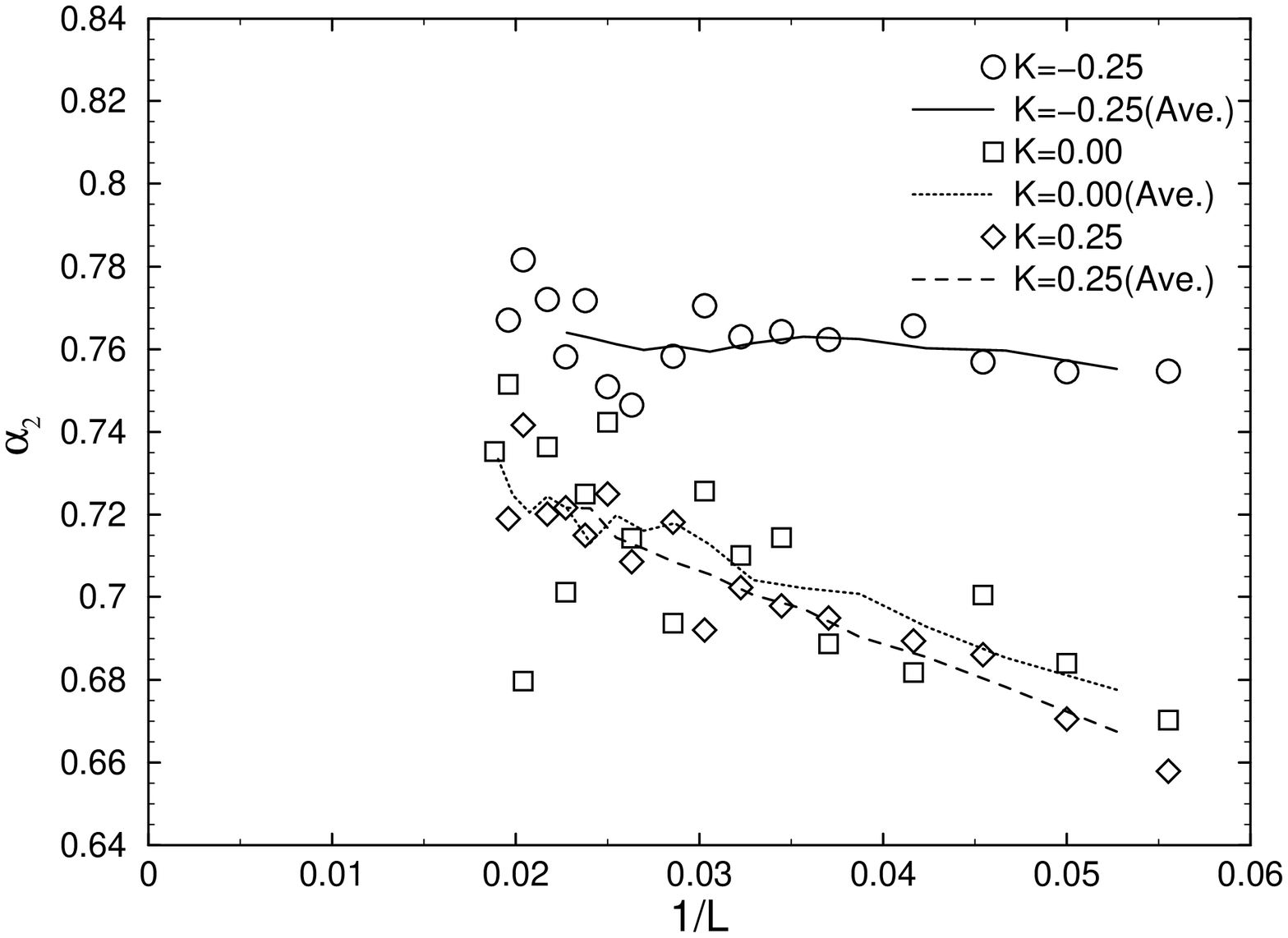}}
\caption{ 
Finite size scaling approximants for the surface roughness exponent  
$\alpha_2$ of the second moment
in the stationary state of the 2D BCSOS model at $K= \pm 0.25$ and 0.}
\label{BCSOS-a2K}
\end{figure}

\begin{figure}
\centerline{\epsfxsize=8cm \epsfbox{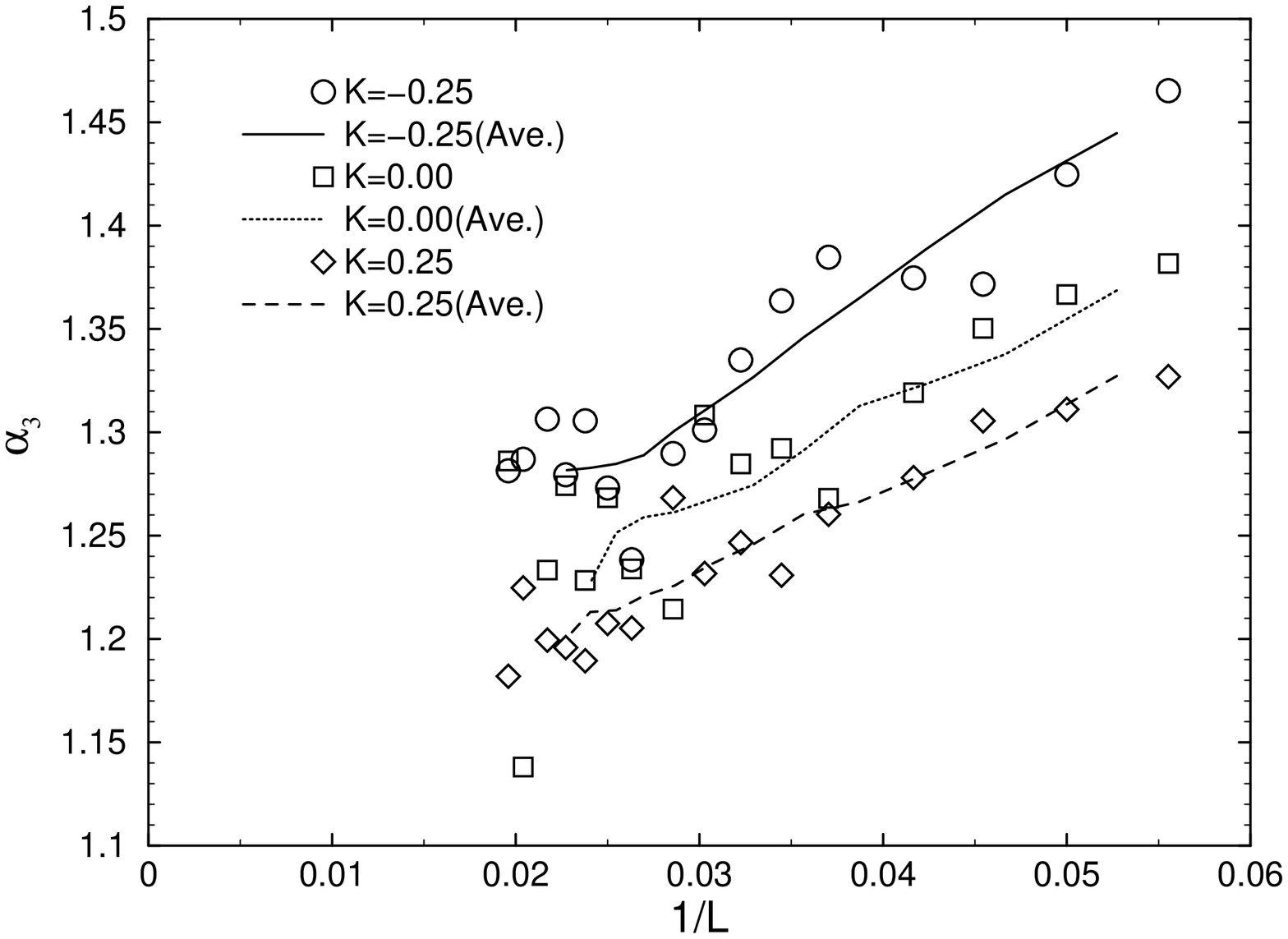}}
\caption{ 
Finite size scaling approximants for the surface roughness exponent  
$\alpha_3$ of the third moment
in the stationary state of the 2D BCSOS model at $K= \pm 0.25$ and 0.}
\label{BCSOS-a3K}
\end{figure}

\begin{figure}
\centerline{\epsfxsize=8cm \epsfbox{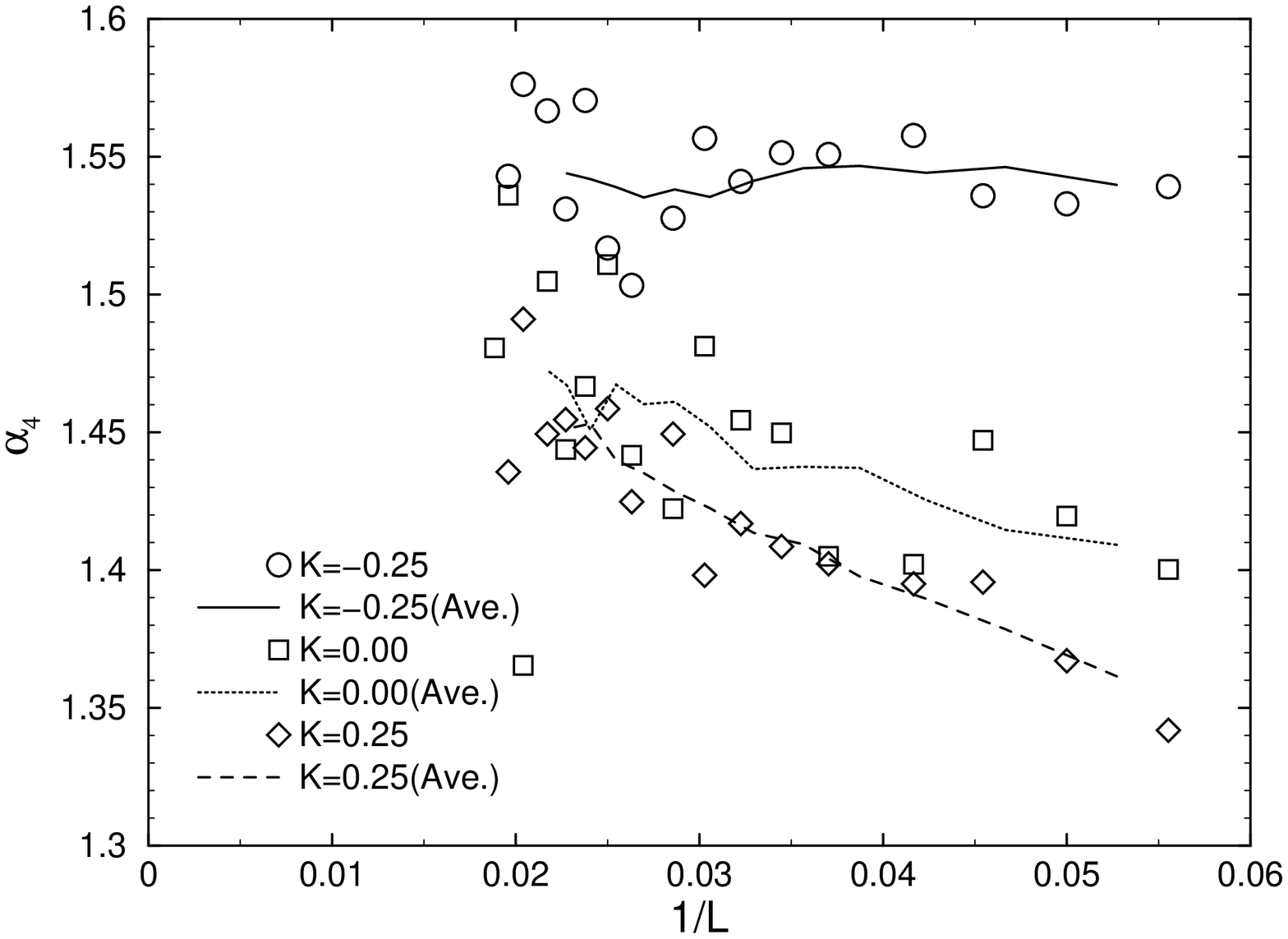}}
\caption{ 
Finite size scaling approximants for the  surface roughness exponent  
$\alpha_4$ of the fourth moment
in the stationary state of the 2D BCSOS model at $K= \pm 0.25$ and 0.}
\label{BCSOS-a4K}
\end{figure}

We performed quantitative MC runs at $K=\pm 0.25, \pm 0.1$
for system sizes up to $L=60$   
(with $2~10^6$ MC step runs after $4~10^3$ initial MC steps).
Figs.~\ref{BCSOS-a2K}, ~\ref{BCSOS-a3K}, and ~\ref{BCSOS-a4K}
show the $\alpha_n(L)$ approximants as function of $1/L$ at $K=\pm 0.25$. 
The $K=0$ data is included as reference.
The corrections to scaling in the even moments increase with $K$, and 
are an order of magnitude smaller at $K=-0.25$.
The three curves are consistent with convergence 
towards the same values for $\alpha_2$ and $\alpha_4$.
Naively these values point to an $\alpha$ smaller than $\frac{2}{5}$,
but still consistent with the KKL value if the crossover scaling exponent is large
(see section \ref{sec:3}).
The corrections to scaling in $\alpha_3$ are less clear cut.
At first glance the three curves seem more or less parallel
(which would suggest a continuous variation in $\alpha_3$ with $K$), 
but they actually converge to indistinguishable values for $\alpha_3$ 
given the error bar from the numerical MC noise.

\begin{figure}
\centerline{\epsfxsize=8cm \epsfbox{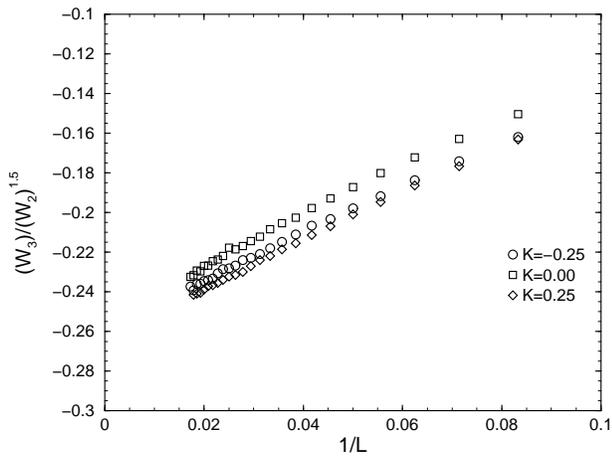}}
\caption{ 
Finite size scaling behaviour of the skewness amplitude ratio $R_3=W_3/W_2^{1.5}$ 
in the stationary state of the 2D BCSOS model at $K= \pm 0.25$, and 0.}
\label{BCSOS-R3K}
\end{figure}

\begin{figure}
\centerline{\epsfxsize=8cm \epsfbox{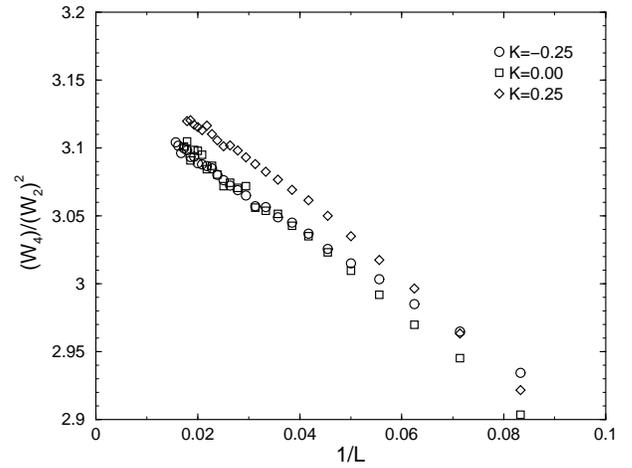}}
\caption{ 
Finite size scaling behaviour of the fourth moment amplitude ratio $W_4/W_2^2$ 
in the stationary state of the 2D BCSOS model at $K= \pm 0.25$  and 0.}
\label{BCSOS-R4K}
\end{figure}

The amplitude ratio's $R_3=W_3/W_2^{3/2}$ and
$R_4=W_4/W_2^{2}$ are shown in Figs.~\ref{BCSOS-R3K} and ~\ref{BCSOS-R4K}.
Like before these amplitude ratio's are more stable than the numerical 
values of $\alpha_n$.
The data confirm universal $K$-independent values of $R_n$
(although $R_4$ drifts off by a few percent at $K=0.25$).

\begin{figure}
\centerline{\epsfxsize=8cm \epsfbox{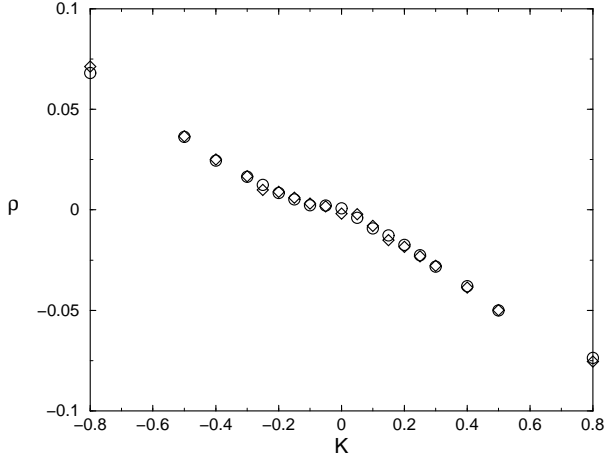}}
\caption{ 
The difference in the expectation value of local sharp hill tops and 
of local sharp valley bottoms as function of $K$ in the 2D BCSOS model
along 1D cross sections through the surface.
The diamonds (circles) are for lattice size $L=36$ (64).}
\label{BCSOS-1DHV}
\end{figure}

\begin{figure}
\centerline{\epsfxsize=8cm \epsfbox{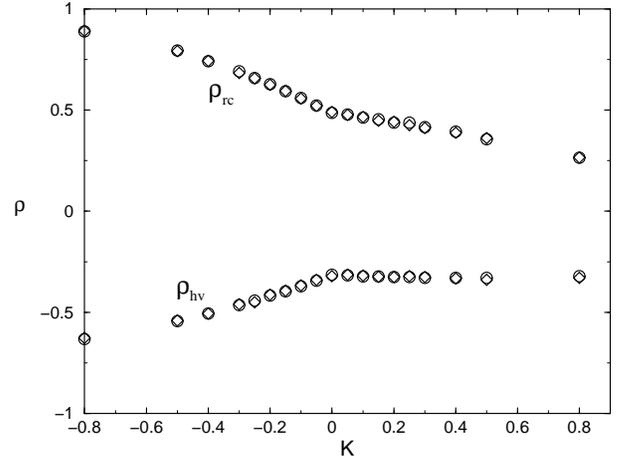}}
\caption{ 
The difference in the expectation value of sharp hill tops and 
of sharp valley bottoms, $\rho_{hv}$, 
and local sharp ridges and canyons, $\rho_{rc}$, 
as function of $K$ in the 2D BCSOS model.
The diamonds (circles) are for lattice size $L=36$ (64).}
\label{BCSOS-2DHV}
\end{figure}

\begin{figure}
\centerline{\epsfxsize=8cm \epsfbox{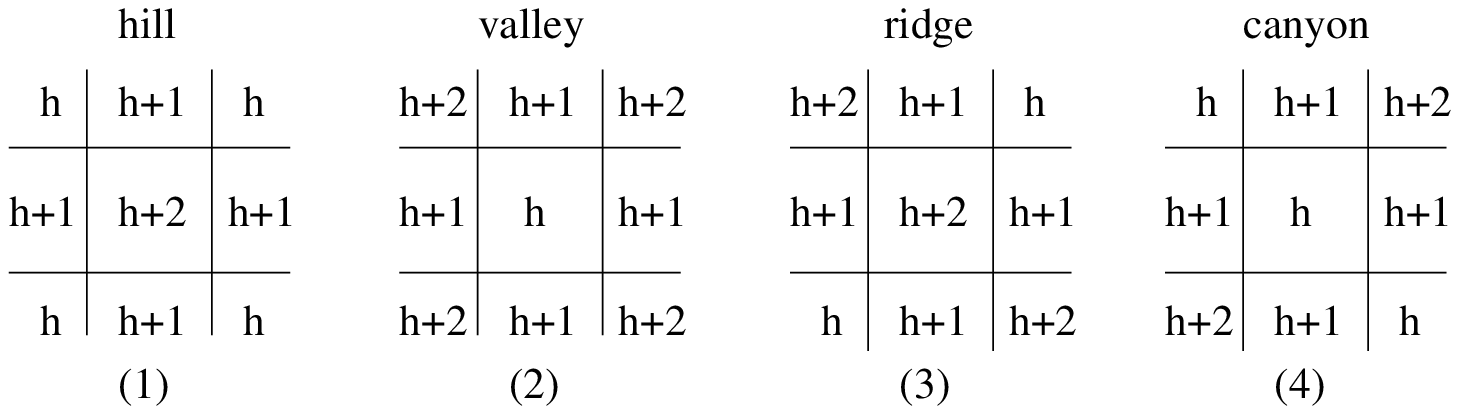}}
\vskip 30 true pt
\caption{ 
Definition of local sharp hill tops, valley bottoms, ridges, and canyons
in the 2D BCSOS model.}
\label{BCSOS-HVRC}
\end{figure}

In $1D$ the skewness amplitude is tunable and 
directly related to the amount of particle-hole symmetry breaking 
at the microscopic level. 
Figs.~\ref{BCSOS-1DHV} and ~\ref{BCSOS-2DHV}
show how the microscopic particle-hole 
asymmetry varies as function of $K$ in the $2D$ BCSOS model.
We measure several local quantities.
Fig.~\ref{BCSOS-1DHV} shows the
density difference ratio, $\rho= (\rho_h-\rho_v)/(\rho_h+\rho_v)$,  
between local sharp hill tops, $\rho_h$,  and local sharp valleys, $\rho_v$, 
as seen along $1D$ cross-sections of the crystal.
Fig.~\ref{BCSOS-2DHV} shows 
the density difference $\rho_{hv}$ between local sharp hill tops and local sharp valleys bottoms 
in the 2D surface  and also the difference between
sharp ridges versus sharp canyons $\rho_{rc}$
as defined in Fig.~\ref{BCSOS-HVRC} .
All three quantities vary dramatically with $K$.
This demonstrates that the local skewness coupling constant 
$u_{sk}$ varies significantly with $K$. 
Sufficiently to expect a large variation in $R_3$ 
if  (macroscopic) skewness is tunable. 
Since this not the case, $R_3\simeq -0.27$ is most likely a universal property 
of the stationary state.

The curves in Fig.~\ref{BCSOS-1DHV} and Fig.~\ref{BCSOS-2DHV} have kinks at $K=0$.
These are caused by the definition of the deposition probabilities, 
$p= {\rm min}(1, \exp(-\Delta E))$.
The probabilities are temperature dependent for some configurations  
but constant, $p=1$, for others.
At $K=0$ they are being reshuffled.
The curves of Fig.~\ref{BCSOS-Ka2} have similar dips at 
$K=0$ for small $L$, but these vanish with system size.
This is  another illustration of the insensitivity of the macroscopic
length scale properties of the stationary state on the 
local properties.
 
\section{Conclusions}\label{sec:6}

The most important result of this paper is that 
the amplitude ratio's $R_3=W_3/W_2^{1.5}$ and $R_4=W_4/W_2^{2}$ 
of the stationary state moments
converge to the same value in the KK and BCSOS model, 
$R_3= -0.27 (1)$ and $R_4= +3.15 (2)$.
At the start of this project we expected that the skewness, $R_3$, 
would be the most sensitive parameter to test the universality of the KPZ
stationary state properties in $2D$.
However, $R_3$ is numerically much more stable than the 
surface roughness critical exponent $\alpha$.
$R_3$ takes the same value in the KK and BCSOS models, 
and does not vary significantly in the BCSOS model with the temperature parameter $K$;
in contrast to the  strong variation in the microscopic measures of particle-hole 
asymmetry.
The $2D$ KPZ stationary state skewness is universal,
unlike $1D$ where it is tunable.

The differences in the numerical values for $\alpha$ in the KK and BCSOS model
have been a puzzle for a long time. 
The corrections to FSS scaling in the KK model are small, compared to
the MC noise and point clearly to the KKL value $\alpha=\frac{2}{5}$.
The FSS corrections to scaling in the BCSOS model are large.
The stationary state roughness exponent 
converges naively but systematically to a smaller value $\alpha\simeq 0.38$.
However, our  FSS analysis
shows that the data is  consistent with  $\alpha=\frac{2}{5}$ 
if the leading corrections to scaling exponents are dominated by 
$y_{ir}\simeq -0.6~(2)$ for the even moments and 
$y_{ir}\simeq -1.7~(3)$ for the odd ones. 
These values are different, but in the same range as 
predicted by simple minded power counting.
This is an important issue, 
because different values of $\alpha$ in the KK and BCSOS models,
and variations with $K$ in the latter,
would imply non-universality of $2D$ KPZ scaling,
and open up the possibility of e.g., a continuously varying $\alpha$.
Our data suggest one single KPZ fixed point with unique exponents.
It is most likely L\"assig's stationary state.
  
This work is supported by NSF grant DMR-9700430.

\narrowtext
\raggedcolumns

\vspace{2cm}

\pagebreak

\pagebreak

\pagebreak

\end{multicols}

\begin{references}
\bibitem{KPZ}
M.~Kardar, G.~Parisi, and Y-C.~Zhang,Phys.~Rev.~Lett. {\bf 56}, 889 (1986).

\bibitem{rev1} 
J.~Krug and H.~Spohn in {\it Solids Far from Equilibrium: Growth,
Morphology and Defects}, ed. C. Godr\`eche (Cambridge University Press,
Cambridge, 1991).

\bibitem{rev2} 
P.~Meakin, Physics Reports {\bf 235}, 189 (1993).

\bibitem{rev3} 
J.~Krug in {\it Scale Invariance, Interfaces, and Non\--Equilibrium
Dynamics},ed. A.~McKane, M.~Droz, J.~Vannimenus, and D.~Wolf 
(Plenum, NY, 1995).

\bibitem{rev4}  
T.J.~Halpin-Healy and Y.C.~Zhang, Physics Reports {\bf 254}, 215 (1995).

\bibitem{KPZ-mf} 
M.~L\"assig and H.~Kinzelbach, Phys.~Rev.~Lett. {\bf 78}, 903 (1997); 
K.~Wiese, Phys.~Rev.~E {\bf 56}, 5013 (1997);
C.~Castellano, M.~Marsili, L.~Pietronero, Phys.~Rev.~Lett. {\bf 80}, 4830 (1998).

\bibitem{Dhar} 
D.~Dhar, Phase Transitions {\bf 9}, 51 (1987).

\bibitem{Spohn}
L-H.~Gwa and H.~Spohn, Phys.~Rev.~Lett.~{\bf 68}, 725 (1992);
and, Phys.~Rev.~A {\bf 46}, 844 (1992).

\bibitem{JN-MdN} 
J.~Neergaard and M.~den~Nijs, Phys.~Rev.~Lett. {\bf 74}, 730 (1995).

\bibitem{Derrida} see e.g.,
B.~Derrida, M.R.~ Evans, V.~ Hakim, and V.~ Pasquier,
J.~Phys.~A {\bf 26}, 1493 (1993).


\bibitem{KimKos} 
J.~M.~Kim and J.~M.~Kosterlitz, Phys.~Rev.~Lett.~{\bf 62}, 2289 (1989).

\bibitem{KK-temp} 
J.G.~Amar and F.~Family, Phys.~Rev.~Lett.~{\bf 64}, 543 and 2334 (1990);
J.~Krug and H.~Spohn, Phys.~Rev.~Lett.~{\bf 64}, 2332 (1990); 
J.~Kim, T.~Ala-Nissila, and J.M.Kosterlitz, Phys.~Rev.~Lett.~{\bf 64}, 2333 (1990). 

\bibitem{BCSOS-num} 
For numerical results on the $2D$ BCSOS model see e.g.: 
D.~Liu and M.~Plischke, Phys. Rev. B {\bf 38}, 4781 (188);
M.~Koita and A.C.~Levi, J.Phys.A {\bf 25}, 3121 (1992);
B.M.~Forrest and Lei-Han~ Tang,  Phys.~Rev.~Lett.~{\bf 64}, 1405 (1990);
and the above review papers \cite{rev1,rev2,rev3,rev4}.

\bibitem{Lassig} 
M.~L\"assig, Phys.~Rev.~Lett. {\bf 80}, 2366 (1998). 

\bibitem{JN-MdN-sk}
M.~den Nijs and J.~Neergaard, J.~Phys.~A {\bf 30}, 1935 (1997).

\bibitem{temp-sk}
J.~Krug, P.~Meakin, and T.~Halpin-Healy, Phys.~Rev.~A {\bf 45}, 638 (1992).

\bibitem{MdN-RSOS}
M.~den Nijs, J.~Phys.~A {\bf18} (1985) L549-556.

\bibitem{MdN-PR} 
M.~den Nijs, chapter 4 in {\it The Chemical Physics of 
Solid Surfaces and Heterogeneous Catalysis}, Vol.7,
edited by D.~King, Elsevier (Amsterdam, 1994).

\bibitem{6vertex} 
See e.g., H.~van Beijeren and I.~Nolden, in {\it Structures and dynamics of
Surfaces}, edited by W.~Schommers and P.~von Blanckenhagen  Vol.~2
(Springer, Berlin 1987).

\end{references}
\end{document}